%% file: main.tex
\documentclass[fleqn,10pt]{wlscirep}
\usepackage[utf8]{inputenc}
\usepackage[T1]{fontenc}

\usepackage{graphicx} 
\DeclareGraphicsExtensions{.png,.jpeg} 
\usepackage{cite} 
\usepackage{tabularx} 
\usepackage{amsmath} 
\usepackage{amsfonts} 
\usepackage{subcaption}
\usepackage{wrapfig}
\title{Leveraging Multimodal CycleGAN for the Generation of Anatomically Accurate Synthetic CT Scans from MRIs}

\author[1, *]{Leonardo Crespi}
\author[1]{Samuele Camnasio}
\author[2, 3]{Damiano Dei}
\author[2, 3]{Nicola Lambri}
\author[2, 3]{Pietro Mancosu}
\author[2, 3]{Marta Scorsetti}
\author[1]{Daniele Loiacono}

\affil[1]{Department of Electronics, Information and Bioengineering, Politecnico di Milano, Milan, Italy}
\affil[2]{Department of Biomedical Sciences, Humanitas University, Milan, Italy}
\affil[3]{Radiotherapy and Radiosurgery, IRCCS Humanitas Research Hospital, Rozzano, Italy}

\affil[*]{Corresponding author: leonardo.crespi@polimi.it}

\begin{abstract}
    In many clinical settings, the use of both Computed Tomography (CT) and Magnetic Resonance (MRI) is necessary to pursue a thorough understanding of the patient's anatomy and to plan a suitable therapeutical strategy; this is often the case in MRI-based radiotherapy, where CT is always necessary to prepare the dose delivery, as it provides the essential information about the radiation absorption properties of the tissues. Sometimes, MRI is preferred to contour the target volumes. However, this approach is often not the most efficient, as it is more expensive, time-consuming and, most importantly, stressful for the patients. To overcome this issue, in this work, we analyse the capabilities of different configurations of Deep Learning models to generate synthetic CT scans from MRI, leveraging the power of Generative Adversarial Networks (GANs) and, in particular, the CycleGAN architecture, capable of working in an unsupervised manner and without paired images, which were not available. Several CycleGAN models were trained unsupervised to generate CT scans from different MRI modalities with and without contrast agents. To overcome the problem of not having a ground truth, distribution-based metrics were used to assess the model's performance quantitatively, together with a qualitative evaluation where physicians were asked to differentiate between real and synthetic images to understand how realistic the generated images were. The results show how, depending on the input modalities, the models can have very different performances; however, models with the best quantitative results, according to the distribution-based metrics used, can generate very difficult images to distinguish from the real ones, even for physicians,  demonstrating the approach's potential.
\end{abstract}

\begin{document}

\flushbottom
\maketitle
%
%
\thispagestyle{empty}


\input{sections/intro.tex}
\input{sections/sota.tex}

\input{sections/methods.tex}
\input{sections/results.tex}
\input{sections/conclusions.tex}

\bibliography{ref2}



\section*{Author contributions statement}

LC wrote the paper; 
LC, SC and DL designed and conducted the experiments; 
NL, DD and P.M. helped in performing the experiments and with the analysis of the results; 
DL, P.M. and MS supervised the experiments. All authors reviewed the manuscript.

\section*{Data Availability}
Data from the CHAOS dataset is public and available at the following link:  .

\begin{itemize}
    \item \textbf{CHAOS dataset}: the data that support the findings of this study are openly available in Kaggle at https://chaos.grand-challenge.org/Data/ .
    \item \textbf{AUTOMI dataset}: the data that support the findings of this study are not publicly available.
    \item \textbf{Trained models}: the trained models are available upon request to the corresponding author.
\end{itemize}

Data from AUTOMI is proprietary and will not be publicly available. 



\end{document}

%% file: sections/intro.tex

\section{Introduction}\label{sec:introduction}
Computed Tomography (CT) serves as a widely spread imaging modality for diagnostic purposes across the medical field, as it is one of the most reliable and accurate techniques available, and it allows the visualisation of internal anatomical structures in a non-invasive way precisely. 
Another widely used imaging technique is Magnetic Resonance (MRI), which offers distinct views of the targeted anatomical district, evidencing areas with a high content of hydrogen, such as fat and soft tissues; 
the two techniques are not interchangeable as they are based on radically different physical principles and are used to highlight different tissues and structures, depending on the clinical necessity~\cite{suetens_2009}.

These two tools are only occasionally used side by side, as, typically, there is a specific clinical question to answer that requires only one.
However, scenarios where having both is necessary do exist: it is the case for specific radiotherapy (RT) treatments, like Total Marrow and Lymphoid Irradiation (TMLI)~\cite{wong2020tmi}, where there is the necessity of whole-body contouring, better performed exploiting the rich contrast in MRI images. 
With \emph{contouring}, we refer to the process of identifying and delineating the anatomical structures of interest in the images, which is a crucial step in the RT treatment planning process as it allows to define the target volumes to be irradiated with the dose and the tissues to be spared, thus minimising the risk of damaging healthy tissues and the occurrence of unwanted side effects. 
CT scans are extensively used for this task, bearing the advantage of being necessary for every RT treatment to create the delivery plan, as they contain the specifics about the radiation absorption properties of the patient's tissues, which are fundamental to computing the dose distribution, and seldom offer sufficient information to contour regions of interest (ROI) adequately. 
In some cases, however, with whole-body treatments being a prominent example, radiation oncologists prefer using MRI~\cite{khoo2006new} for the contouring.

An approach with the double examination introduces, however, a series of issues from different standpoints: primarily, for the patients, it means having to undergo two distinct examinations, which adds a lot of stress and discomfort, both physical and psychological; it has to be noted that, in the case of TMLI for instance, the patients in question are already in a very fragile state: generally, they are affected by haematological malignancies, like acute leukaemia or lymphoma, and are undergoing a bone marrow transplant; also, they seldom are towards the more senior side of the population. 
For these reasons, every additional source of stress is not desirable. 
Secondly, performing two scans is more expensive and requires more time, delaying the treatment time because of the time needed to take the exams and process and use the images for the planning. 
Another reason for additional concern in specific scenarios might be the extra dose of ionising radiation provided by the CT scanner; 
however, in the case of TMLI and RT in general, this is not particularly relevant as the dose delivered during the treatments makes the radiation absorbed during the acquisition negligible.
Moreover, the MRI and the CT have to be co-registered, adding a non-trivial step to the process and a possible additional source of error.

From these premises, it can be inferred how a system capable of providing physicians with both CT and MRI from a single acquisition would be a great advantage as it would significantly reduce costs and time, improve the efficiency of the protocols and diminish the stress for the patients. 
Deep Learning (DL) can help in this sense, as with the recent advancements in the field, models based on Convolutional Neural Networks (CNN) have shown great promise in transforming the image domain; in this work, we aimed at studying this task, applied to the transition from MRI to CT, as in the case previously described, the relevant anatomical structures of most significant interest for the use-cases are most evident in MRI images. 
Incidentally, such a tool would be a precious resource to generate accurate and realistic data in a field where data is notoriously relatively scarce and difficult to obtain.
Having a system capable of generating accurate synthetic medical images would make it possible to increase the richness of the dataset of choice, including patients from, possibly, different hospitals and populations, thus reducing the risks of overfitting and bias in the learnt representations~\cite{tommasi2017deeper}.
Also, generated data is less subject to privacy concerns as the synthetic images do not generally carry with them information that would allow to track back the original patient, or they might even be used as harmonisation tools to transform the original data and prevent leakages.

The core objective of this work revolves around leveraging advancements in artificial intelligence and deep learning for image-to-image translation between the two imaging modalities.
The aim is to synthesise CT images from available MRI inputs, with a vision to mitigate the operational and financial burdens on healthcare practitioners and patients and reduce the associated psychological or physical challenges from obtaining dual imaging modalities.
Said images must represent the same anatomical structures, with their geometry, opacity and specific textural properties, to be as medically and anatomically accurate as possible.
Unpaired image-to-image translation architectures have been analysed to select suitable architectures to achieve these objectives.
The focal anatomical district in this work is the abdominal region, encompassing critical organs such as the liver, spleen, kidneys, and the inferior section of the lungs.
This anatomical selection offers a broader relevance compared to other regions; for instance, while significant, studies centred on cerebral images are inherently specific to that region only, bearing less universality with the rest of the body as the variety of tissues is less diverse.
The investigative lens is primarily focused on CycleGAN architectures\cite{zhu2017unpaired}.
This work dissects the generation capabilities of various models trained to yield CT scans from different data sources and connects the generation with the combination of multiple MRI modalities. 
The methodology described encompasses diverse evaluation strategies to gauge model performance, ascertaining the effectiveness of generative models without a paired validation dataset.
This work is carried out in collaboration with medical personnel from the Humanitas Research Hospital, who qualitatively assessed the images generated from a medical perspective.

%% file: sections/sota.tex
\section{Related Works}\label{sec:sota}

DL image generation, not only in medical imaging but in general, has been a very active research field in the last few years.
Many interesting applications are available, especially for artistic and media-related purposes, as they can be highly catchy for the general public.
The use of image generation in medical imaging is, however, mainly related to the generation of synthetic data, which can be used for a variety of purposes, from
data augmentation to create missing modalities in MRI,~\cite{giacomello2019transfer}.
In general, the common denominator for this kind of system and application is the use of Generative Adversarial Network (GAN), which, introduced in 2014 by Ian Goodfellow et al.~\cite{goodfellow2020generative}, constitute a framework employing two distinct models: a generator $G$ and a discriminator $D$, that play a min-max game, respectively trying to generate realistic data and to distinguish between real and generated data.

\begin{wrapfigure}{l}{0.5\columnwidth}
    \centering
    \includegraphics[width=0.4\columnwidth]{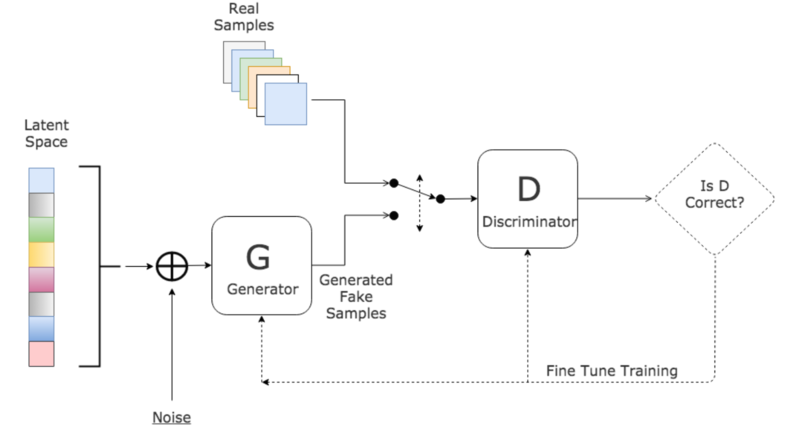}
    \caption{Base scheme of a GAN.~\cite{hitawala2018comparative}}\label{fig:gan}
\end{wrapfigure}

The general basic scheme of a GAN is shown in Fig.~\ref{fig:gan}.
The architecture's potency emanates from the adversarial loss capability to manifest a distribution in the synthetic images analogous to the actual data distribution.
Since GAN's inception, numerous methodologies drawing on its foundational principle have emerged, including the CycleGAN~\cite{zhu2017unpaired}, our principal reference model, which endeavours to produce images indistinguishable from the target dataset.
Several GAN variants with enhanced network architectures and mathematical optimization techniques have been introduced since its first appearence~\cite{uemura2019ensemble,mutasa2020advanced,onishi2020investigation,javaid2018capturing,frid2018gan,gao2020performance,kaur2021diagnosis,calimeri2017biomedical}.
A category of GANs referred to as conditional GANs (CGANs) facilitates conditional image generation from the generator, producing images with specific attributes~\cite{yi2019generative,mirza2014conditional}.
In this sense, both CycleGAN~\cite{zhu2017unpaired} and Pix2Pix~\cite{isola2017image}, among the most influential models, fall under the umbrella of CGANs and cater to image-to-image translation, converting an input image into a synthetic image that meets a specific criterion.
These architectures allow data creation via domain adaptation, implying that the trained models can incorporate adjustments from one data source to another~\cite{sankaranarayanan2018generate}.
Data augmentation is one of the foremost applications of GANs in medical imaging, a notorious field for scarce data availability where this kind of tool is a necessity: several efforts have been made to develop systems to address this particular problem exploiting the capabilities of this family of neural models with particular success targeting classification tasks~\cite{muramatsu2020improving, wang2020combination,xu2019semi,ge2020enlarged,wang2020semi,yu2020synthesis}.
Specifically, Xu et al.~\cite{xu2019semi} introduced a semi-supervised attention-guided CycleGAN for MRI image augmentation aimed at brain tumour classification, capable of producing tumour images from normal ones and vice versa.
Muramatsu et al.~\cite{muramatsu2020improving} can be an illustrative example of CycleGAN utilized to create mammographic masses from lung nodules, aiming at data augmentation for breast mass classification.
Wang et al.~\cite{wang2020semi} ascertained that merging WGAN~\cite{arjovsky2017wasserstein} with progressive growth~\cite{karras2017progressive} outperformed the WGAN with gradient penalty (WGAN-GP)~\cite{gulrajani2017improved} and Pix2Pix GAN~\cite{isola2017image} for pulmonary adenocarcinoma classification; they incorporated a stitch layer into the generator net (dubbed StitchGAN) to merge low-dimensional images into a comprehensive image, simplifying the intricacy of full-size image transformation and generating MRI image pairs for clinically significant classification. 
Ge et al.~\cite{ge2020enlarged} devised a pairwise generative architecture to craft synthetic multimodal MRI images, thus augmenting a multimodal dataset tailored for brain tumour classification.
In another innovative twist, Yu et al.\cite{yu2020synthesis} suggested substituting the conventional CNN with a Capsule network~\cite{sabour2017dynamic} in the GAN's discriminator, enhancing the modelling of the image's hierarchical relationships.
An additional difficulty to image synthesis, particularly relevant in the medical field, is posed when the necessity is to find data augmentation methods for segmentation tasks, as in this case, the synthetic images have to be paired with the corresponding labels, which in this context typically refers to an image where pixels or voxels are associated with a category index, denoting specific relevant objects.
Several recent works have obtained significant results~\cite{shin2018medical,sandfort2019data,jiang2019cross,zhang2019skrgan,chaitanya2019semi,cao2020improving}.
Most of these proposed methodologies are grounded in the domain adaptation concept discussed previously.
For instance, semantic labels of anatomical structures might produce coherent artificial medical images.
When paired with the labels, the resulting images can be incorporated into the training dataset for a segmentation network.
Cao et al.\cite{cao2020improving}, and Shin et al.\cite{shin2018medical} adopted this label image translation technique to produce PET and CT images, as well as atypical brain MRI images, respectively, aiming at tumour segmentation.
Augmented labels, allowing tumour location, shape, and size alterations, were the input for generating synthetic images showcasing diverse anatomical characteristics.
Such cross-domain translation can also be observed between various imaging techniques: Sandfort et al.\cite{sandfort2019data} employed a CycleGAN to convert contrast-enhanced CT images to their non-contrast-enhanced counterparts, targeting organ segmentation.
Conversely, Jiang et al.\cite{jiang2019cross} orchestrated pseudo-MRI image synthesis from CT images, regulated by a pair of concurrently trained GANs, focusing on lung tumour segmentation within MRI. 
These papers have preserved structural correspondence between synthetic images and labels derived from the original image sources, a key aspect. 
A comprehensive analysis by Skandarani et al.~\cite{skandarani2023gans} evaluates the performance of GANs trained on medical datasets for generating synthetic images that remarkably resemble original images.
This analysis juxtaposes various GAN architectures, shedding light on the most effective ones for image generation and convincingly demonstrating the potential of these generated images to deceive human experts in a visual Turing test.
However, this work focuses on image translation, specifically synthesizing a synthetic CT scan corresponding to a provided MRI.
For tasks of image-to-image translation, an exemplary methodology was presented by Armanious et al. ~.\cite{armanious2020medgan}, who utilized a GAN architecture amalgamated with non-adversarial losses, facilitating the translation of PET to CT for brain imagery with promising results. 
The MedGAN framework found applicability in an unsupervised setting, as demonstrated in another study by the same authors~\cite{armanious2019unsupervised}.
A dual experiment incorporating paired and unpaired data for the translation of brain CT to MR was introduced by Cheng-Bin Jin et al.\cite{jin2019deep} through the MR-GAN architecture.
Other explorations in image generation entail the generation of absent modalities, as exemplified in the study by Alogna et al.\cite{alogna2020brain}.
This research utilized multi-input pix2pix and GAN architectures to generate absent MRI brain image modalities, specifically with T1-weighted scans (both with and without contrast enhancement) and T2-weighted alongside T2-fluid-attenuated inversion recovery scans.
Another noteworthy endeavour by Gao et al.\cite{gao2021generating} aimed at generating CT scans from sub-par beam CTs, employing three distinct architectures, pix2pix, CycleGAN, and AGGAN, to assess the efficacy in this high-definition image generation task.
Studies have also focused solely on unpaired data generation, especially concerning brain tumour scans.
A pertinent example would be the investigation by Jelmer M. Wolterink et al.\cite{wolterink2017deep}, which employed the classic CycleGAN architecture introduced by Zhu et al.~\cite{zhu2017unpaired} to synthesize CT scans using sagittal MRI scans of the human brain as input.
This study compares the performance of the CycleGAN in this particular task with traditional paired image-to-image translation techniques, highlighting instances where models trained on unpaired data surpassed their paired counterparts in performance.

%% file: sections/methods.tex

\section{Methodology}\label{sec:Models_Description}

\subsection*{Models}
\begin{figure}
	\begin{subfigure}{0.48\columnwidth}
		\centering
		\includegraphics[width=\columnwidth]{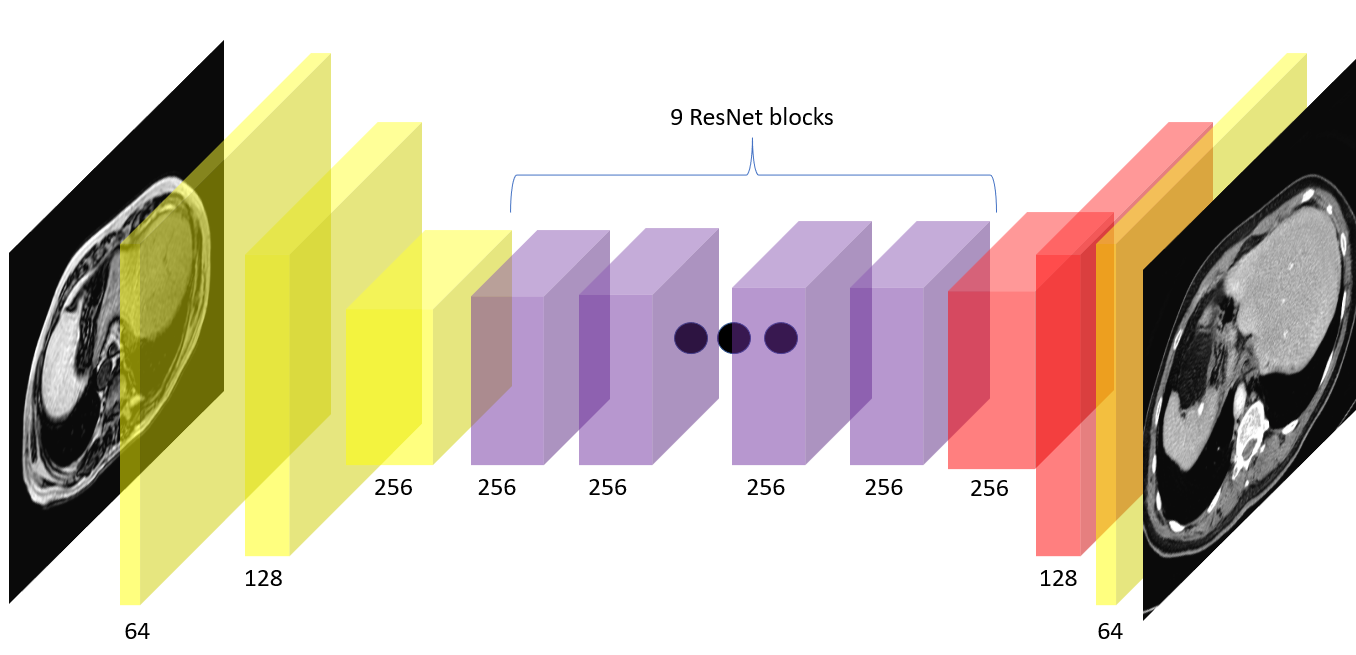}
		\caption{Single-input generator (forward and backward cycle).}\label{fig:GenCycleGAN}
	\end{subfigure}%
	\hspace*{\fill}
		\begin{subfigure}{0.48\columnwidth}
		\centering
  		\includegraphics[width=\columnwidth]{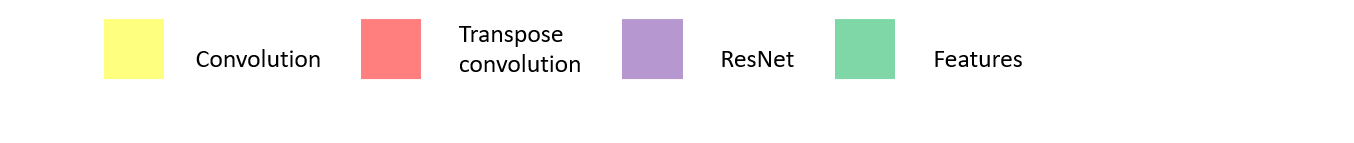}
		\includegraphics[width=\columnwidth]{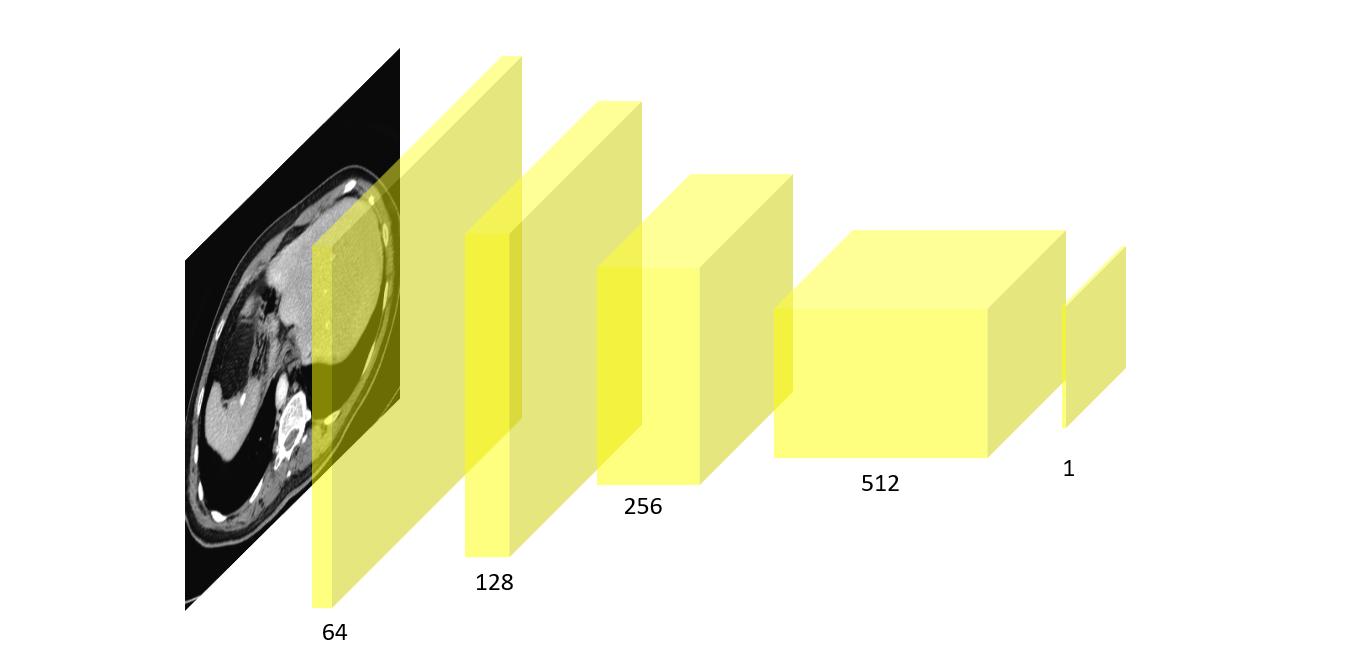}
		\caption{Discriminator.}\label{fig:discCycleGAN}
	\end{subfigure}
	\begin{subfigure}{0.48\columnwidth}
		\includegraphics[width=\columnwidth]{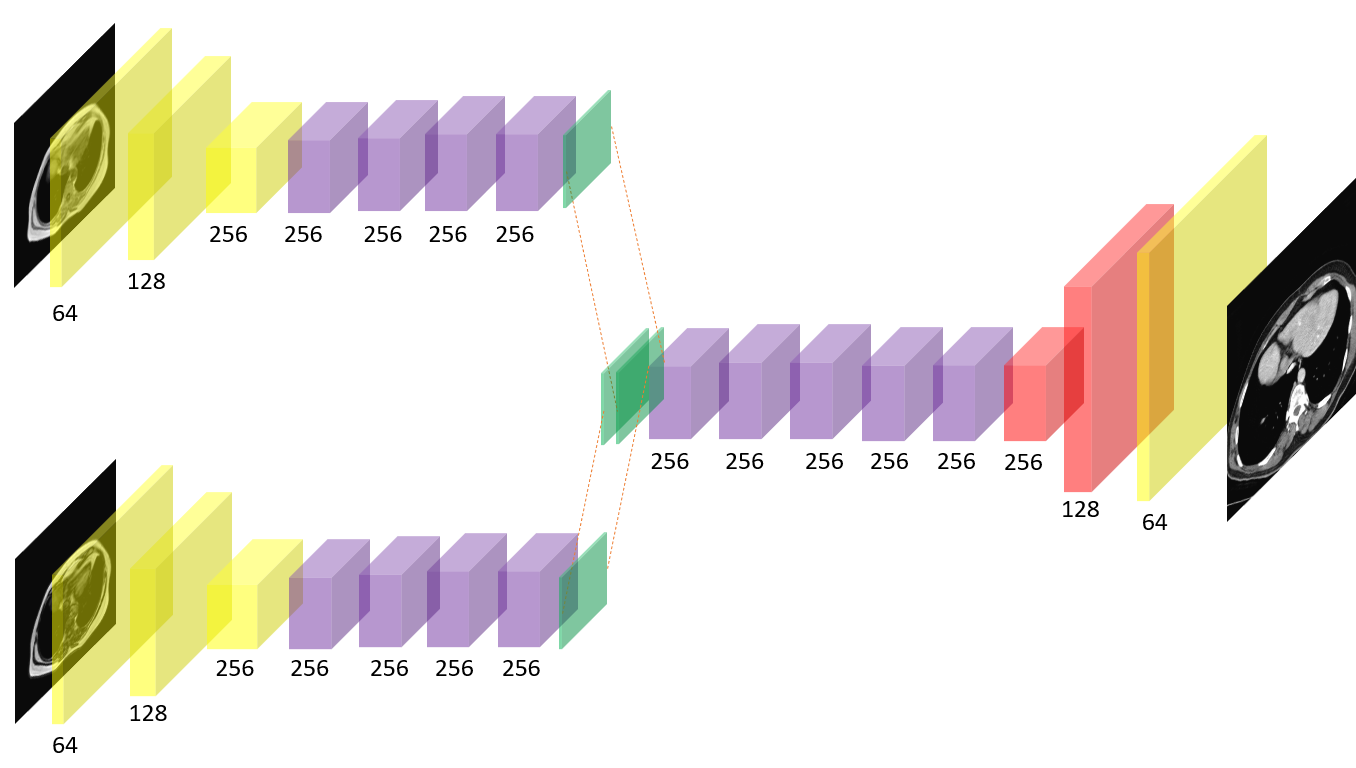}
		\caption{Multimodal generator (forward cycle).}\label{fig:legendaForwardNet}
	\end{subfigure}%
	\hspace*{\fill}
	\begin{subfigure}{0.48\columnwidth}
		\includegraphics[width=\columnwidth]{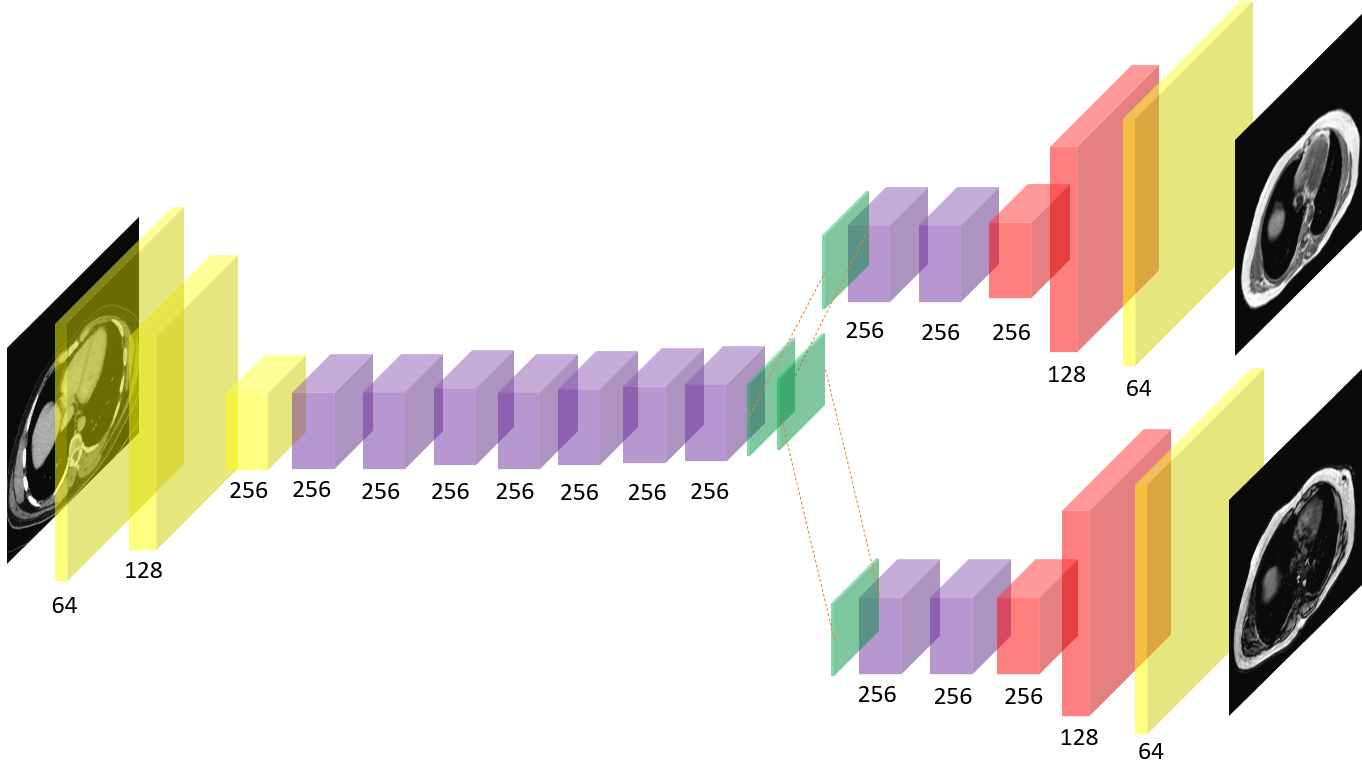}
		\caption{Multimodal generator (backward cycle).}\label{fig:legendaBackwardNet}
	\end{subfigure}
	\caption{Schematic representation of the CNN used in this work.}
\end{figure}

The models presented in this work are based on the CycleGAN architecture~\cite{zhu2017unpaired} and make use of different combinations of input, resulting in single input and multimodal networks, depending on the number of images taken in input; the input is always 2D, using single slices of both MRI and CTs and keeping the complexity lower than with a three-dimensional approach, as sometimes in medical imaging this allows for better outcomes~\cite{crespi2022are}. 
Regarding the implementation, the generator architecture is inspired by Johnson et al.'s design~\cite{johnson2016perceptual}. It incorporates nine residual blocks sandwiched between two downsampling and two upsampling blocks, as shown in figure~\ref{fig:GenCycleGAN}.
The downsampling blocks are built with, in sequence: \emph{convolution}, \emph{ normalization}, and \emph{activation (ReLU)}.
Similarly, after the residual blocks, there are two upsampling blocks with the same scheme except for using \emph{transpose convolutions} instead of regular convolutions, aimed at increasing the size of the layer's input. 

The discriminator draws its architecture from the PatchGAN design, detailed in the Pix2Pix paper~\cite{isola2017image}.
This approach assesses the authenticity of $70\times70$ overlapping patches within the source image.
Such a design permits a reduced parameter set in comparison to a full-image discriminator and offers the flexibility to process images without concerns about their size;
a high-level scheme is shown in figure~\ref{fig:discCycleGAN}.
Two types of CycleGAN networks were used in this study: single-input and multimodal: the former uses a single slice as an input, while the latter takes two corresponding slices from two different modalities of the same slice.
Often, T1 MRIs are acquired in both in-phase (IP) and out-of-phase (OOP) modalities, and it is possible to use both of them as input for the model, as their differences can help distinguish fat from tissues with similar properties~\cite{suetens_2009}.
The same cannot be said for T2 MRI, as they have different purposes than T1-weighted ones, and it is less frequent that they are both acquired. Therefore, developing a system using both would be more challenging to apply in a clinical setting.
The multimodal CycleGAN is configured to accept 2 or 3 inputs for the MRI-to-CT generator, with the reverse transformation producing the corresponding 2 or 3 outputs.
Consequently, there is a dedicated discriminator for the CT image and one for each MRI modality.
In the forward generator, a single branch is used for each input modality to extract features, which are subsequently merged into a latent space for output CT generation~\ref{fig:legendaForwardNet}.
Conversely, the backward generator begins with a singular input that transitions to its specific latent space, facilitating the generation of multiple MRI modalities as output to keep cycle consistency.
The feature extraction block at the end of the input branches is organized as one convolution block, followed by two downsampling blocks and four residual blocks.

Several setups have been tested, varying inputs and modalities, to understand how the model behaves using different modalities:
\begin{itemize}
    \item \textbf{All-in}: single input model, trained with both in-phase and out-of-phase T1 MRI and contrast-enhanced CT, all data from the Chaos Dataset;
    \item \textbf{In-phase}: single input model, trained with in-phase only MRI and contrast-enhanced CT, all data from Chaos Dataset;
    \item \textbf{In-phase+}: single input model trained with T1-weighted MRI from the Chaos Dataset plus some from the Cancer Imaging Archive to test a more varied dataset, CT are contrast-enhanced from the Chaos dataset;
    \item \textbf{T2}: single input model with T2 weighted MRI and contrast-enhanced CT from the Chaos dataset;
    \item \textbf{In 'n out}: 2-input model, with correlated in-phase and out-of-phase T1 MRI images and contrast-enhanced CT, all images from Chaos Dataset;
    \item \textbf{In 'n out HUM}: 2-input model, in-phase and out-of-phase MRI from Chaos Dataset and non-contrast-enhanced CT from the Humanitas Dataset;
    \item \textbf{Triple}: 3-input model, with correlated T1(in-phase and out-of-phase) and T2 MRI images and contrast-enhanced CT, all images from Chaos Dataset.
\end{itemize}

\subsection*{Datasets}
Three sets of images have been used in this study:
\begin{itemize}
    \item \textbf{Chaos CT}: CT scans of the abdominal region from the CHAOS challenge~\cite{kavur2020comparison, CHAOS2021}; 40 scans from different patients are contained, from potential liver donors with no tumours, lesions or other pathologies; the orientation and alignment are the same for all the images, which consist of 16-bit DICOM files with a resolution of $512\times 512$, x-y spacing between 0.7 and 0.8mm, and 3 to 3.2 mm inter-slice axial distance, corresponding to an average of 90 slices per scan with a total of 1367 images; all the scans are contrast-enhanced. 
    \item \textbf{Chaos MRI}: MRI scans of the same region from the CHAOS challenge; it contains 40 patients for which two sequences were acquired; T1-Dual, which is a combination of in-phase and out-of-phase T1-weighted images, and T2; 12-bit DICOM images with $256\times 256$ resolution are contained in the set; the inter-slice axial distance varies between 5.5 and 9mm, x-y spacing is between 1.36 and 1.89mm and the average number of slice is 36, with a total of 1594 slices. They are not paired with the CT scans or from the same patients, making it impossible to use a supervised method as no ground truth is available.
    \item \textbf{AUTOMI CT}: full body CT scans from the Humanitas Research Hospital in Milan; 100 patients are contained in DICOM files with 16-bit depth and $512\times 512$ resolution, $3-5mm$ axial distance and $1-2mm$ pixel spacing; the scans are not contrast-enhanced and are not paired with any MRI; the abdominal region has been cropped from the whole body scans to match the other sets, resulting in 5970 images. These sets were acquired from patients undergoing TMLI or similar treatments. 
\end{itemize}

CT sets are used independently so that models are trained either on contrast-enhanced images or non-contrast-enhanced ones, but not both together as they are incompatible, and the capability of generating both would imply introducing a conditioning to the network, adding further complexity to the task. 
Conversely, the same MRI set is used for all the experiments, and the desired modalities for each model are selected. 
It must also be noted that contrast-enhanced CT scans are not generally used in RT, and in general, models generating this kind of CT scans would not have broad practical uses in a clinical setting, but they remain an interesting benchmark. 

\subsection*{Evaluation}\label{subsec:metrics-synthetic}
The lack of ground truth to compare the results calls for distribution-based metrics to assess, from a statistical point of view, the \emph{similarity} of the generated images with the real ones. 
Three metrics have been selected for this purpose, among the most popular in literature: Fréchet Inception Distance (FID)~\cite{heusel2017gans}, Kullback-Leibler Divergence (KL) and histogram comparison.

\subsubsection*{Fréchet Inception Distance}
It was introduced specifically for the evaluation of generative models~\cite{heusel2017gans}, GANs in particular, as it gives an estimate of how \emph{distant} the generated images are from the real ones, using a features space derived from a pre-trained CNN, normally the InceptionV3 (hence the name, \emph{Inception}), stripped of its last layers and only used as a features extractor. 
The resulting distributions are then compared using the Frechet distance, also called the Wasserstein-2 distance~\cite{arjovsky2017wasserstein}.
The lower the score, the closer the two distributions and, hence, the more similar the images.
Notably, FID doesn't merely quantify the image's distribution but aims to assess image similarity akin to human perception.
The FID is computed as:
\begin{equation}
d^2((m, C), (m_w, C_w)) = \| m - m_w \|_2^2 + \text{Tr}(C + C_w - 2(CC_w)^{1/2})
\end{equation}
Here, $ (m, C) $ represents the mean and covariance of the Gaussian distribution derived from the actual image set, while $ (m_w, C_w) $ corresponds to the generated dataset.

\subsubsection*{Kullback-Leibler Divergence}
The Kullback–Leibler (KL) divergence can quantify the dissimilarity between two probability distributions.
In its approximated form~\cite{hershey2007approximating}, it computes the difference between the cross-entropy of distributions $p(x)$ and $q(x)$, and the entropy of $p(x)$, represented, in this work, respectively the distribution of real CT images and the synthetic one.
The formulation used in this work is the following:
\begin{equation}
    D_{KL}(p, q) = H(p, q) - H(p) = \int p(x) \log \left( \frac{p(x)}{q(x)} \right) dx =
    \mathbb{E}_p \left[ \log \left( \frac{p(x)}{q(x)} \right) \right]
\end{equation}
A low KL divergence score implies greater similarity between the compared distributions.
Distributions are derived by averaging image histograms, producing an average detail of the pixels' distribution across the sets.

\subsubsection*{Histogram comparison}
Conceptually similar to the KL divergence, histogram comparison has been used to scrutinize the distribution of the generated images.
The methodology for distribution computation mirrors that of the KL divergence, and the histogram is compared according to two metrics:

\begin{itemize}
    \item{Correlation}:
        \begin{equation}
            d(H1,H2) = \frac{\sum_I (H1(I) - \bar{H1})(H2(I) - \bar{H2})}{\sqrt{\sum_I (H1(I) - \bar{H1})^2 \sum_I (H2(I) - \bar{H2})^2}}
        \end{equation}
        where:
        \begin{equation}
            \bar{H}_k = \frac{1}{N} \sum_J H_k(J)
        \end{equation}
        and $ N $ stands for the histogram bin count.

    \item{Intersection}:
        \begin{equation}
            d(H1,H2) = \sum_I \min(H1(I), H2(I))
        \end{equation}
\end{itemize}

For these metrics, high scores indicate a higher similarity.
Specifically, \emph{Correlation} peaks at 1.0 for flawless matching, while \emph{Intersection}, being the cumulative minimum value across bins, theoretically ranges from $[0, \inf)$.
The formulations for the metrics are adapted from the official OpenCV documentation~\cite{bradski2000opencv}.

\subsubsection*{Spectral Analysis}
An additional method used to assess the quality of generated images is spectral analysis, which is used to compare the frequency content of the images.
The inspiration was drawn from the work of Frank et al.\cite{frank2020leveraging}, which put forth the idea of training a neural network to differentiate between genuine and synthetically generated images from various network designs by scrutinizing their spectrum.
This strategy hinged on the hypothesis that the spectrum of synthetic images may manifest anomalies, primarily because of the upsampling and downsampling operations that transpire during the image-to-image transformation process; for example, transposed convolution layers, often used in GANs, are known to introduce checkerboard-shaped artefacts~\cite{odena2016deconvolution}.
In this work, instead of training a neural network, images have been converted into their spectral representation, and the distance between generated and original ones has been computed through their correlation. 

\subsubsection*{Qualitative blind assessment with medical doctors}

\begin{figure}
	\begin{subfigure}{0.5\columnwidth}
		\centering
		\includegraphics[width=\columnwidth]{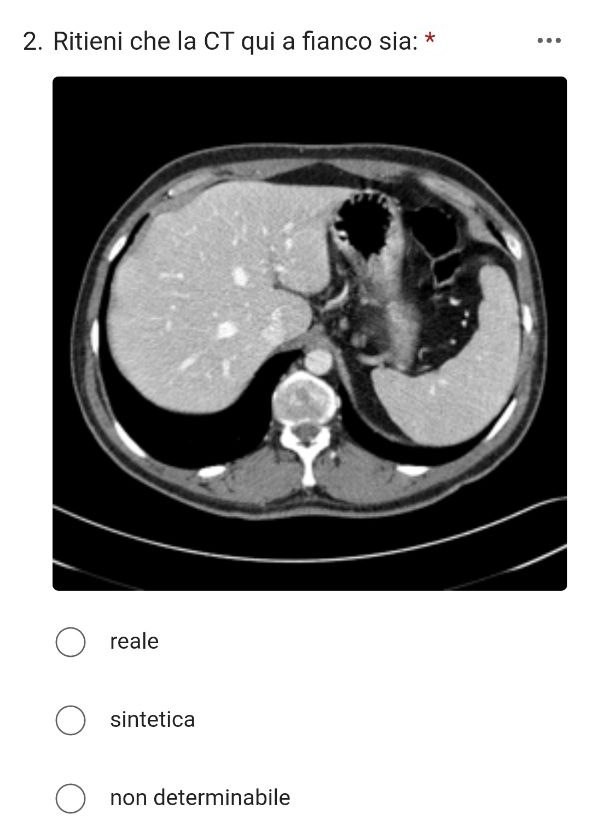}
	\end{subfigure}%
	\begin{subfigure}{0.5\columnwidth}
		\centering
		\includegraphics[width=\columnwidth]{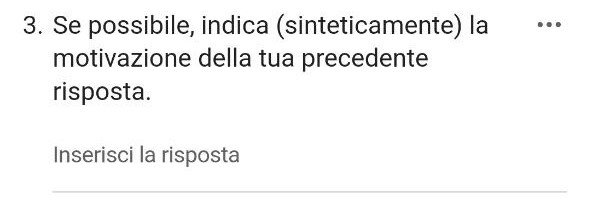}
	\end{subfigure}

	\caption{An example of a portion of the survey proposed to the doctors. The language is \emph{Italian}, as the participants were all Italians.
		\emph{Question 2.} translates to: \emph{"Do you think the image is : -real; -synthetic; -indeterminable"}.
		\emph{Question 3.} translates to: \emph{"If possible, briefly state the motivation for your answer"}.}\label{fig:survey2}
\end{figure}

A qualitative assessment of the generated images has been carried out by medical personnel from the Humanitas Research Hospital. 
The study was conducted by several radiation oncologists, who were asked to evaluate the images and distinguish the real ones from the model's outputs through surveys presenting some randomly selected images in random order, balanced between real and generated. 
Optionally, they were also invited to state the rationale behind each decision, pointing out artefacts, textures, structures, and elements that influenced their judgment.
This approach helped us to understand potential pitfalls in the trained models and furnished insights into the overall quality of the generated scans.
An example of the survey can be seen in~\ref{fig:survey2}.

%% file: sections/results.tex

\section{Results}
\subsection{Quantitative Methods}
One of the most pressing challenges in this work is the lack of paired data, as distribution-based metrics serve more as an estimate of the actual quality of the generated images; even if the distribution is the same, artefacts and structures may have unrealistic shapes or present wrongful intensity patterns while still performing well metrics-wise.

Two subsets of the two real CT datasets, contrast and non-contrast-enhanced from Chaos and AUTOMI, have been kept apart to be compared with the other real images to compute a baseline representing the score obtained from comparing sets of real images.
The scores of our models are then normalized by dividing by their baseline, showing the ratio between the baseline's score and the models'.
This allows for a better comparison of models trained on different kinds of images.
\emph{Baseline} refers to the CHAOS CT baseline, while \emph{hum-baseline} refers to the Automi CT baseline.
All the scores obtained from models trained with contrast-enhanced CT scans are normalized by dividing by the \emph{baseline} value of their layer. Also, the averaged value is divided by the average value of its \emph{baseline}.
The non-contrast-enhanced model is divided by the \emph{hum-baseline}.

\begin{figure}
    \centering
    \includegraphics[width=\columnwidth]{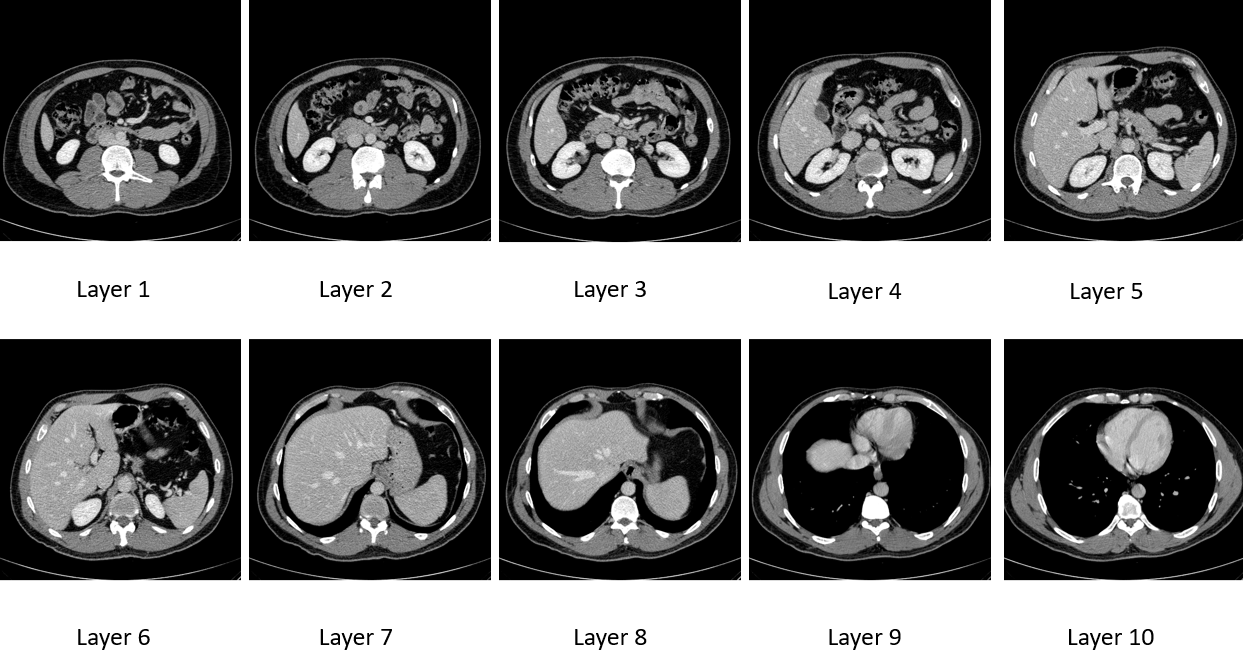}
    \caption{Representative slices for each of the ten portions in which the region is split for the analyses. The splits go from the lower regions encompassing the lowest portion of the kidneys and the bowels to the lowest portion of the lungs in layers 8, 9, and 10.}\label{fig:CTlayers}
\end{figure}

Given the variability from the upper to the lower anatomical regions of the abdomen and the inconsistency in volumes' acquisitions starting or ending at possibly different heights, the anatomical distribution may be substantially different between the two compared datasets;
consequently, images were categorized into axial layers depending on their position along the axial axis.
Ten layers were defined based on the size of observable organs (liver, spleen, kidneys, and lungs) to harmonize the anatomical differences, enhanced by the different sources and settings and usually particularly evident at the borders;
The split aims to consider anatomical subsections of the body which contain similar structures and compare them against each other while having a more profound and finer understanding of the models' performances throughout the volumes. 
For instance, the topmost layers show parts of the lungs, while the bottom layers reach the lower abdominal organs, capturing the appearance of the liver or spleen.
Representative images from each layer, highlighting inter-layer differences, can be seen in figure~\ref{fig:CTlayers}.
Metrics were computed on images within the same layer, after which an average value was calculated.
This method provided insights into the model's performance across different organs across layers, pinpointing which anatomical structures the generative models excelled or faltered with more substantial variability than the central ones. A worsening in scores can be expected.

\subsubsection*{FID}

\begin{figure}
	\centering
	\includegraphics[width=0.7\columnwidth]{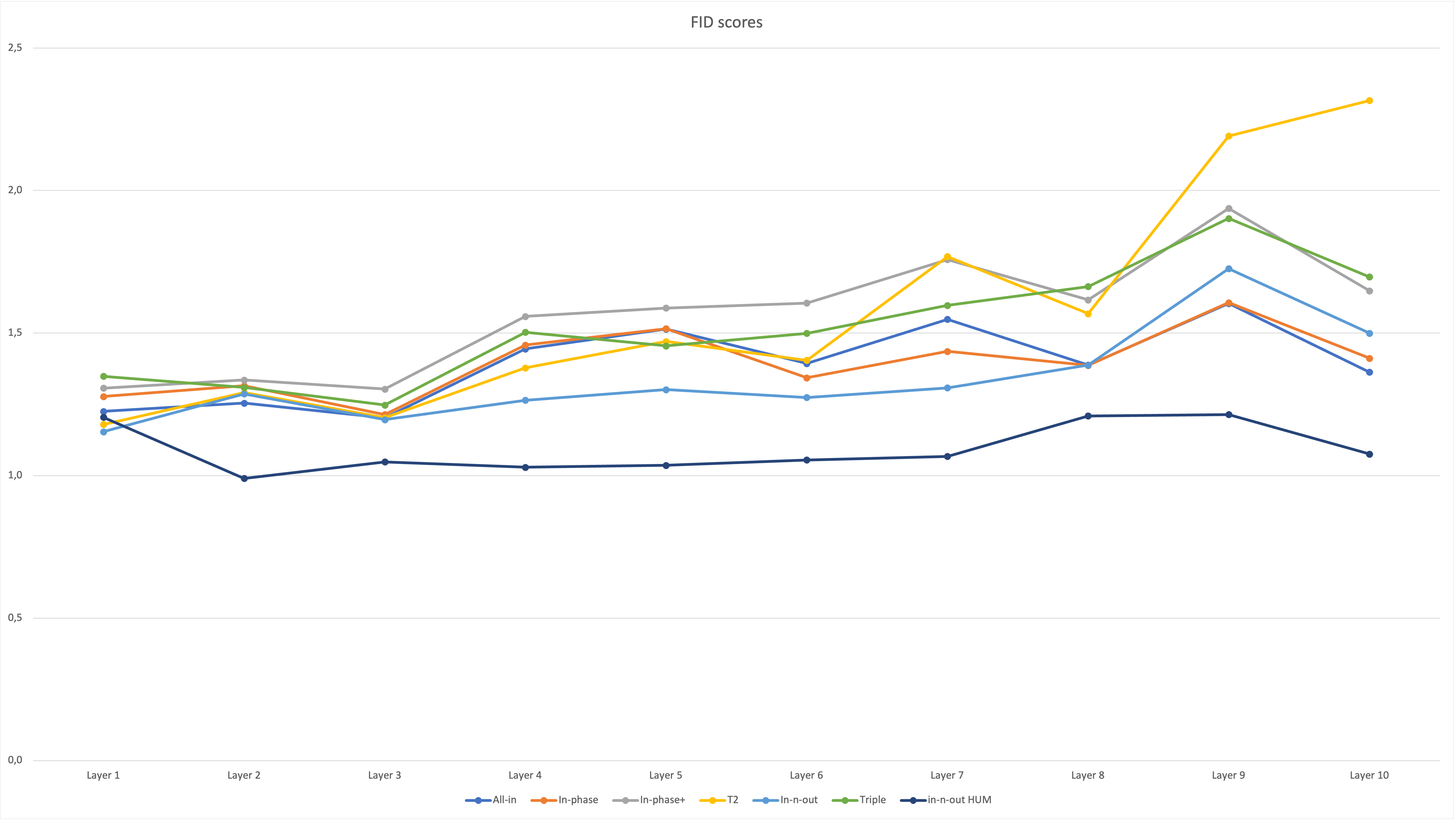}
	\caption{Layer-wise FID scores for each model.}\label{fig:fid_scores}
\end{figure}

The detailed and average scores of the FID are presented in fig.~\ref{fig:fid_scores}.
The findings indicate that the superior scores are derived from the multi-input CycleGAN when utilizing in-phase and out-of-phase T1-weighted MRI images.
This consistency is evident in models trained with contrast-enhanced and non-contrast-enhanced CT, with the former aligning closely with its baseline (\emph{hum-baseline}).
A decline in performance can be noted with the addition of T2-weighted images as the MRI input, which is somewhat unexpected.
A possible reason might be the subtle misalignment of T2 images with corresponding T1 scans, attributed to different acquisition frequencies, which introduces ambiguity and compromises the outcome.
Among the single-input models, the model using T2 images underperformed, hinting at the relative inefficacy of T2-weighted images as a source for synthetic CT. 
The darker values of T2 images, compared to T1, potentially render the contrast of the scans less favourable than their counterpart to generate accurate distributions.
Other models demonstrated commendable outcomes, notably the \emph{All-in} and \emph{In-phase} models.
Their results are closely aligned with the \emph{In-n-out} score.
However, performance deteriorated with varied image types, as observed in the \emph{In-phase+} model.

A pattern in the results showcases subpar outcomes in layers 9 and 10, corresponding to the lung sections, wherein numerous flawed images were produced.
The initial layers, representing the abdomen's lower segments with a glimpse of the liver, also posed challenges, as evidenced by its score contrasting with the scores from the central ones.
The compromised results in the abdomen's extremities could be attributed to limited data availability for these sections or their inherent variability compared to other zones.
For instance, the lung images, often depicted in black, analogous to air or water, or sometimes displaying minimal liver portions, seem particularly difficult.

\subsubsection*{KL Divergence}

\begin{figure}
	\begin{subfigure}{0.5\columnwidth}
		\includegraphics[width=\textwidth]{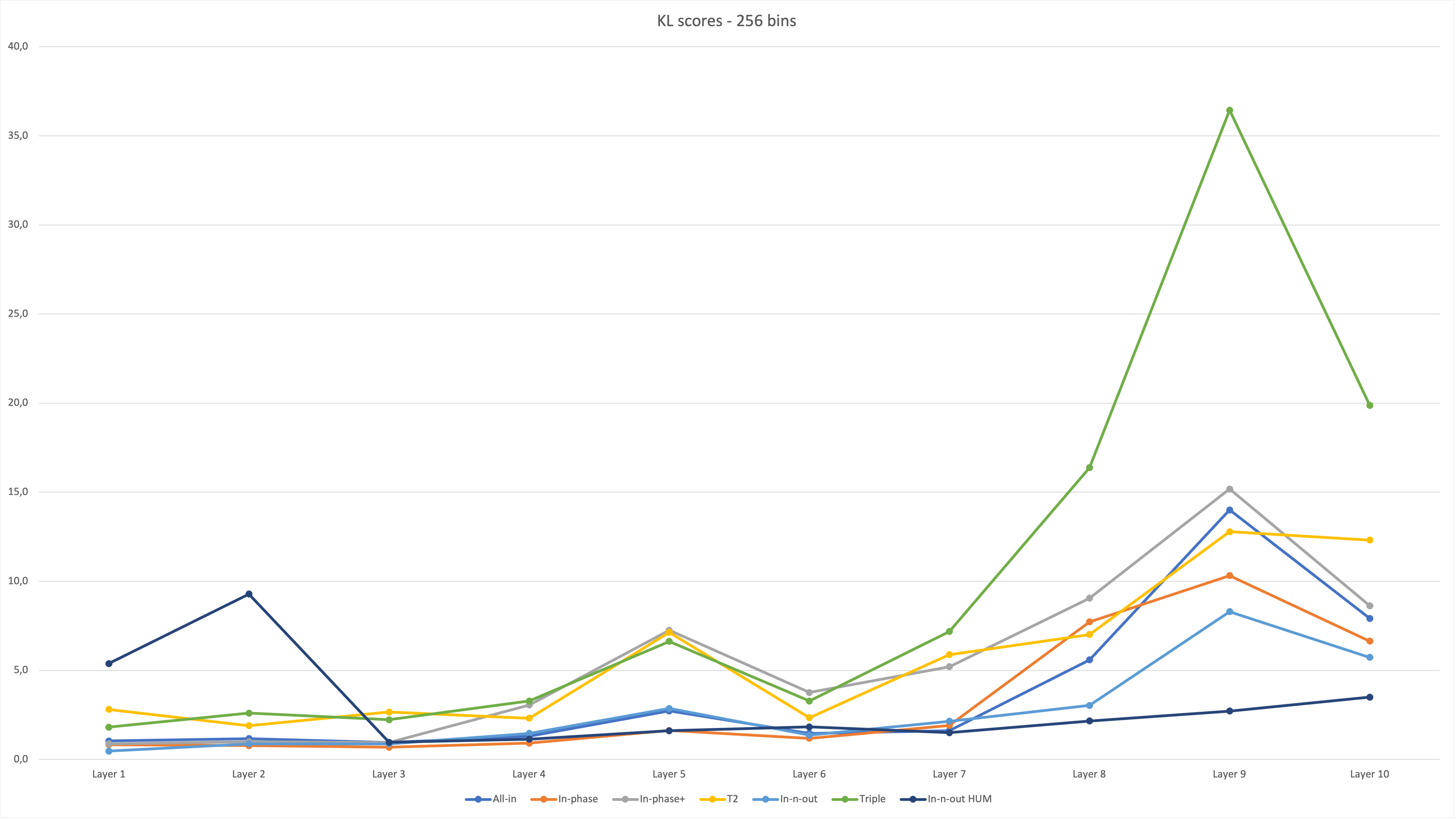}
	\end{subfigure}%
	\begin{subfigure}{0.5\columnwidth}
		\includegraphics[width=\textwidth]{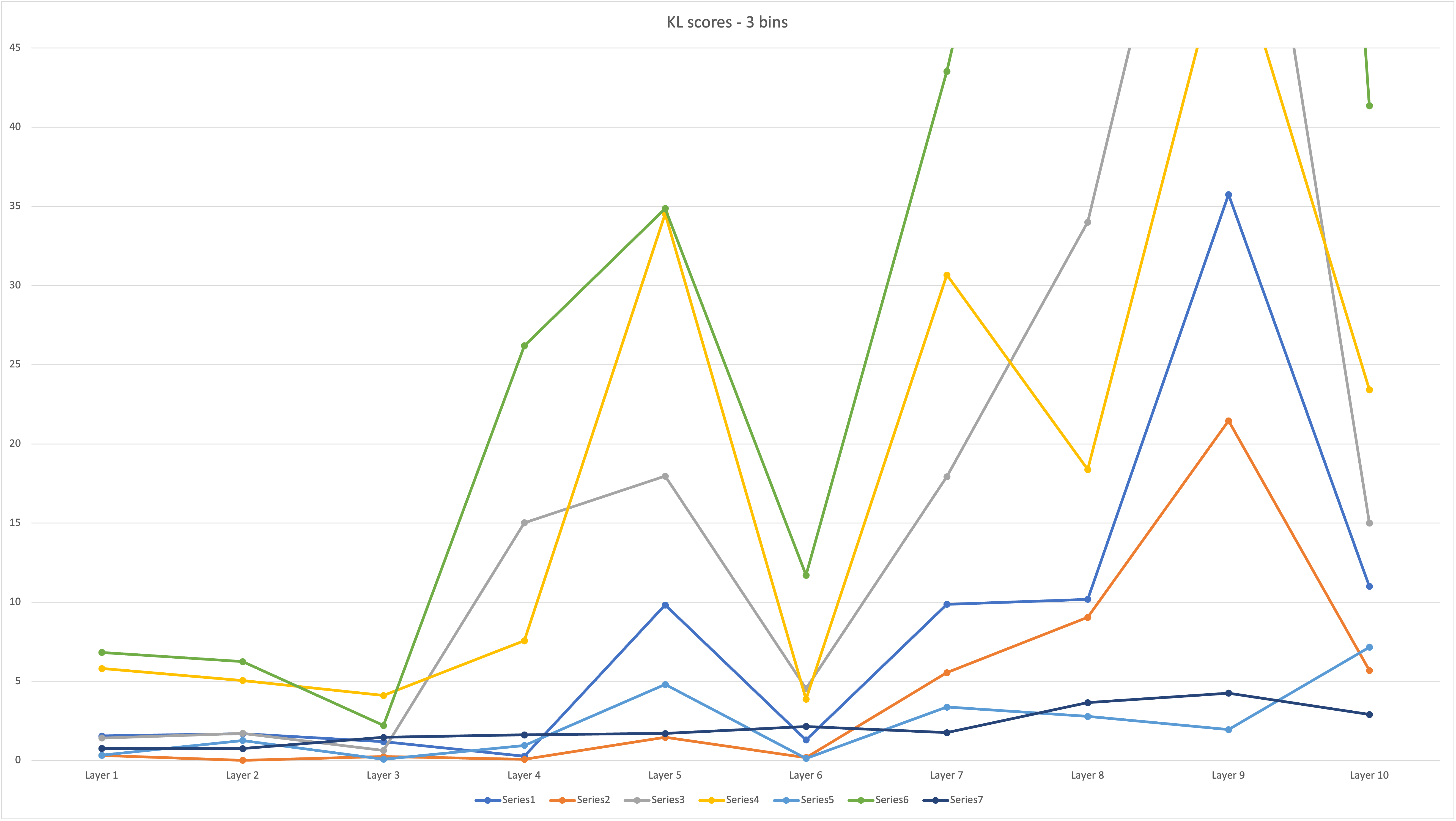}
	\end{subfigure}
	\caption{Layer-wise KL scores for each model; left: 256 bins; right: 3 bins.}\label{fig:kl_scores}
\end{figure}

Two analyses have been performed with the KL divergence metric.
Scores were derived from images processed as both 256-bin and 3-bin histograms.
The 256-bin histograms were employed for grayscale value comparisons to assess the pixel-wise distribution considering all the possible values and, thus, the real distribution itself.
The 3-bins approach is suitable for value comparison related to internal body structure classifications, clustering them according to their radio-opacity: gases and liquids (first bin and darkest values), soft tissues and organs (second bin), and bones (third bin and brightest values);
in this way, it is easier to assess if the overall types of structures are maintained in the synthetic images and if the tissue distribution is coherent, but the analysis is rougher.
The results are shown in fig.~\ref{fig:kl_scores}.
In alignment with the FID explanation, the 256-bins methodology suggests that the \emph{In-n-out} and \emph{In-n-out HUM} models exhibit the best performance as they marginally surpass the \emph{In-phase} and \emph{All-in} models.
The \emph{In-phase+} model presents acceptable outcomes, whereas the \emph{T2} model yields slightly inferior results, and the \emph{Triple} model appears to be underperforming.

The three bins model reveals the same ranking, with the difference that the \emph{In-phase} model has a very similar score to the one obtained by the \emph{In-n-out} model, which is understandable considering the rougher comparison.
Once again, layers in the middle show closer results to the baseline's distribution, demonstrating consistencies in the generation capabilities where the anatomical structures are more consistent.
However, the models' performance is significantly lower in the external layers, especially towards the lungs.

\subsubsection*{Histogram comparison}

\begin{figure}
	\begin{subfigure}{0.5\columnwidth}
		\centering
		\includegraphics[width=\columnwidth]{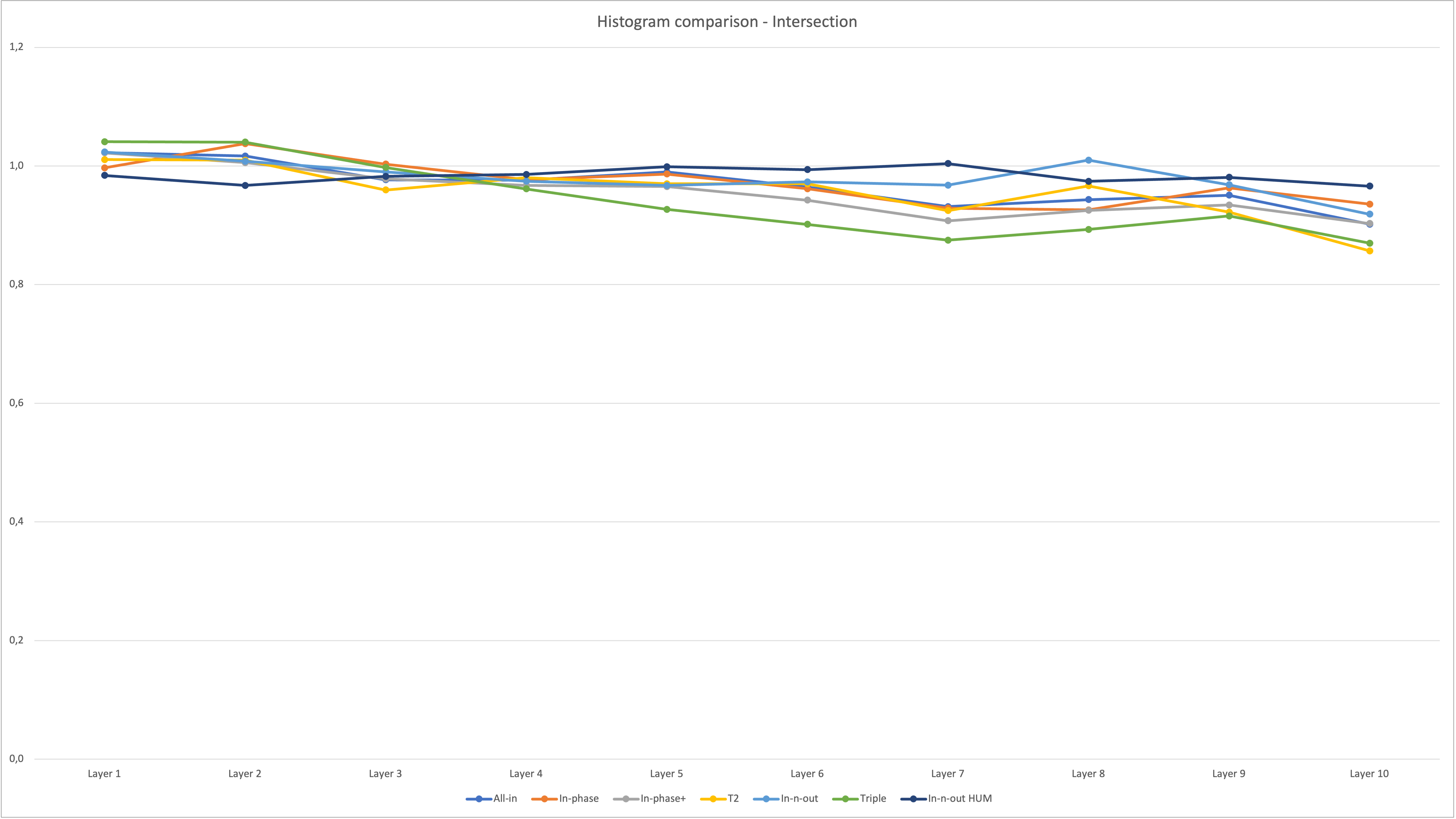}
	\end{subfigure}%
	\begin{subfigure}{0.5\columnwidth}
		\centering
		\includegraphics[width=\columnwidth]{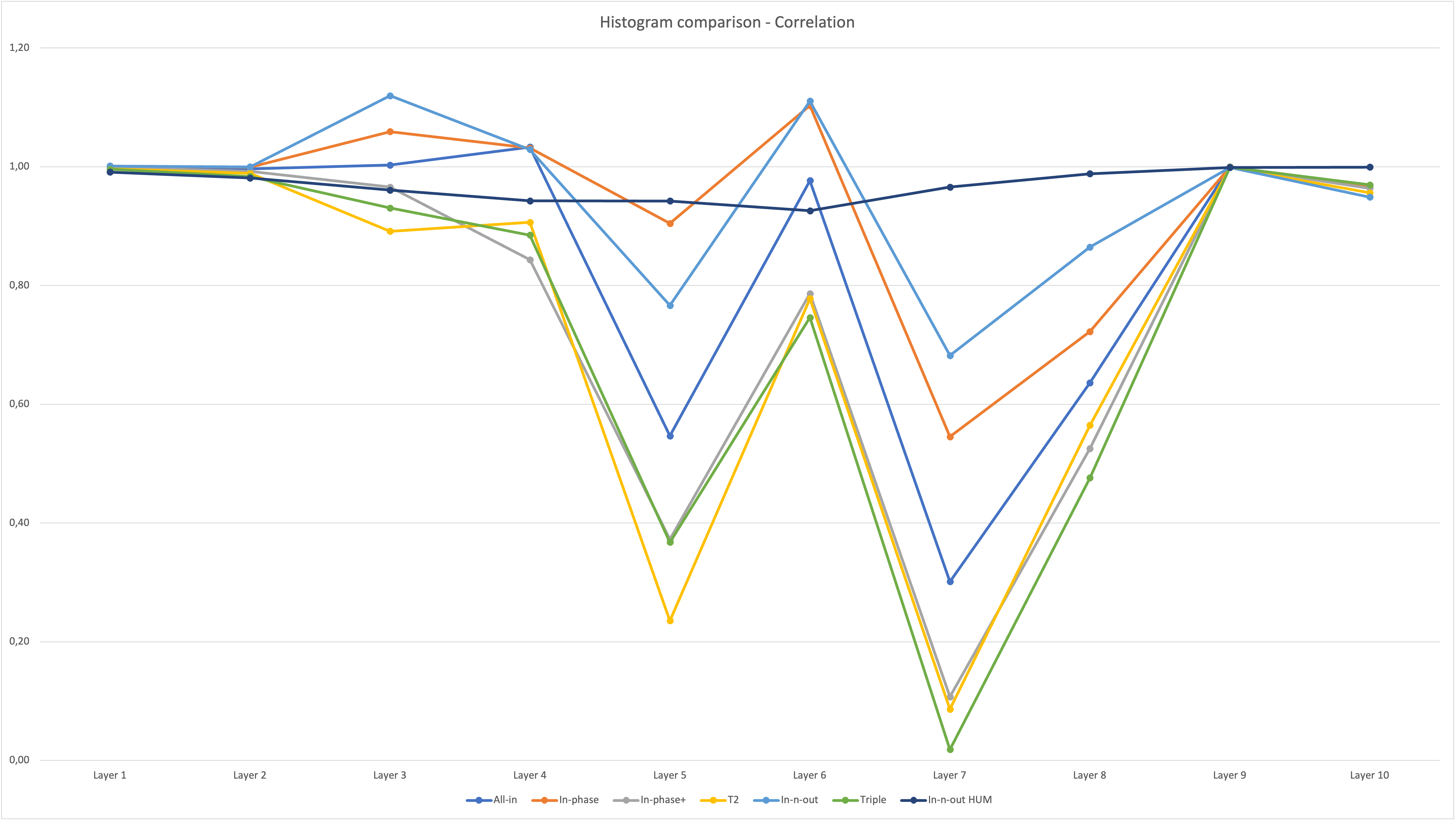}
	\end{subfigure}
    \caption{Layer-wise histogram comparison scores for each model with the metric: left: intersection; right: correlation.}\label{fig:gen_hist}
\end{figure}

Histogram comparison has been performed with the two methods of intersection and correlation, as described in section~\ref{subsec:metrics-synthetic}.
The plots with the scores are shown in fig.~\ref{fig:gen_hist}.
Scores with intersection metric, in this case, show comparable performance across all the settings, with the \emph{In-n-out} and \emph{In-n-out HUM} models standing above the other in the central and rostral layers while being outshined by the others in the caudal ones, which is surprising considering the other metrics. 
However, considering the correlation metric, there is a better result from the \emph{In-n-out HUM} alone, which can keep a robust performance along the axial plane. 
In contrast, all the other models have particularly oscillating scores, even in the central portions.
Once again, \emph{Triple} and \emph{T2} show the worst results, closely followed, this time, by \emph{In-phase+}.
An evident performance worsening in the external layers is not noticeable, but all the models seem to follow the same trend, except for the \emph{In-n-out HUM}. 

\subsubsection*{Spectral analysis}\label{res:spectral_analysis}%

\begin{table}
	\centering
	\begin{tabular}{|p{8em}  c |}
		\hline
		                      & \textbf{Score} \\
		\hline \hline
		\textbf{In-n-out}     & 1,21434        \\
		\textbf{Triple}       & 1,09408        \\
		\textbf{All-in}       & 1,10132        \\
		\textbf{In-phase}     & 1,03135        \\
		\textbf{In-phase+}    & 1,03283        \\
		\textbf{T2}           & 0,94746        \\
		\textbf{In-n-out HUM} & 0,98592        \\
		\hline
	\end{tabular}
	\caption{Average scores obtained from the correlation between spectra of real and synthetic images.}\label{table:spectre}
\end{table}

The outcomes were derived across different layers, akin to the methodology used in the quantitative analysis, and their mean was ascertained and presented in table~\ref{table:spectre}.
The score was computed for all the trained models; in the same way as before, we incorporated two baselines for non-contrast-enhanced and contrast-enhanced CT scans.
The scores obtained were then normalized with respect to their baselines.

\begin{figure}
	\begin{subfigure}{0.45\columnwidth}
		\includegraphics[width=\columnwidth]{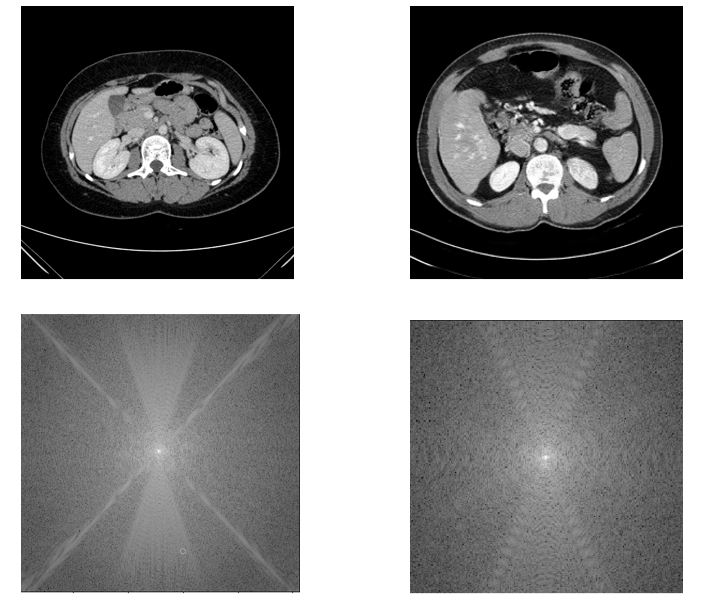}
		\caption{Real vs similar slice from \textbf{In-n-out}, evidencing artefacts and noise in the spectral image of the second.}\label{fig:spectre1}
	\end{subfigure}%
	\begin{subfigure}{0.45\columnwidth}
		\includegraphics[width=\columnwidth]{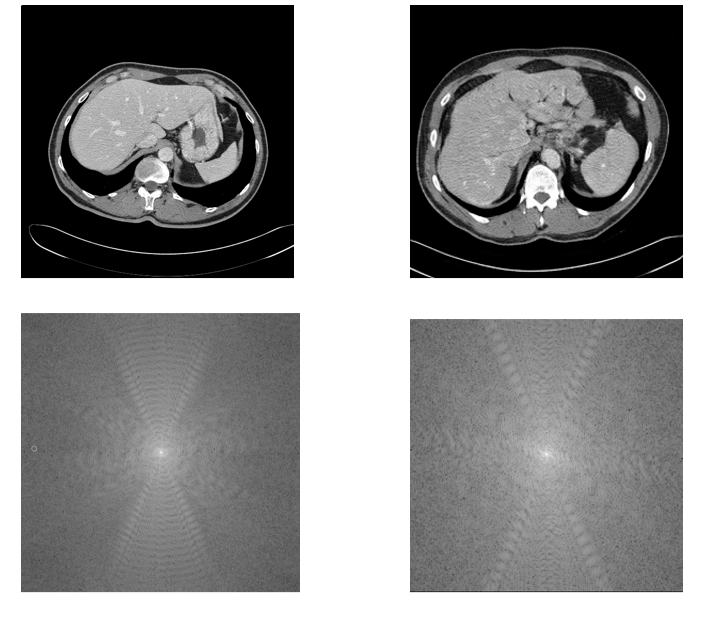}
		\caption{Real vs \textbf{In-n-out} similar output image; the spectra are similar but more noisy.}\label{fig:spectre2}
	\end{subfigure}
	\centering
	\begin{subfigure}{0.5\columnwidth}
		\centering
		\includegraphics[width=\columnwidth]{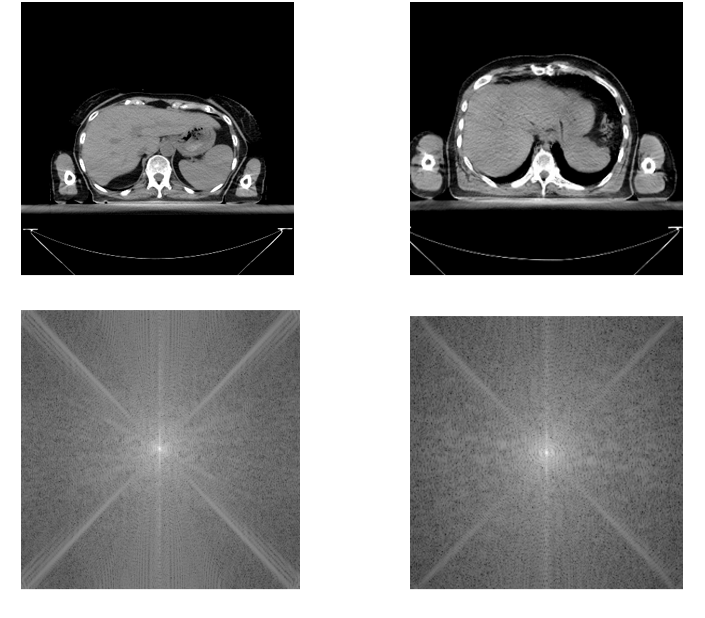}
		\caption{Real vs \emph{In-and-Out HUM}; once again, the noise is pretty evident in the spectrum. }\label{fig:spectre3}
	\end{subfigure}
	\caption{Examples of spectral comparisons between a real CT (left) and a similar synthetic image from the same district (right).}\label{fig:spectre}

\end{figure}

The computed scores do not show any significant distinction in image quality between genuine and synthesized images.
However, a qualitative observation of the spectra can highlight something more accessible for a human observer to understand. By presenting specific analogous CT scans alongside their spectral counterparts, it is evident how the spectrum of generated images generally appears to be more noisy and with less defined patterns, even if the images are similar and show the same structures.
As an illustrative example, fig.~\ref{fig:spectre1} displays a pair of images where the quality of the synthetic one pales in comparison to the authentic scan.
Conversely, fig.~\ref{fig:spectre2} and~\ref{fig:spectre3} feature real and synthetic images with remarkably similar spectra, the same patterns, and much noise in the synthetic ones. 
The source of these noise-like artefacts can be understood by zooming in on the images to closely observe the texture, which is generally way less smooth and precise than in real images.

\subsection{Qualitative assessment}
\begin{figure}

	\begin{subfigure}{0.48\columnwidth}
		\centering
		\includegraphics[width=\columnwidth]{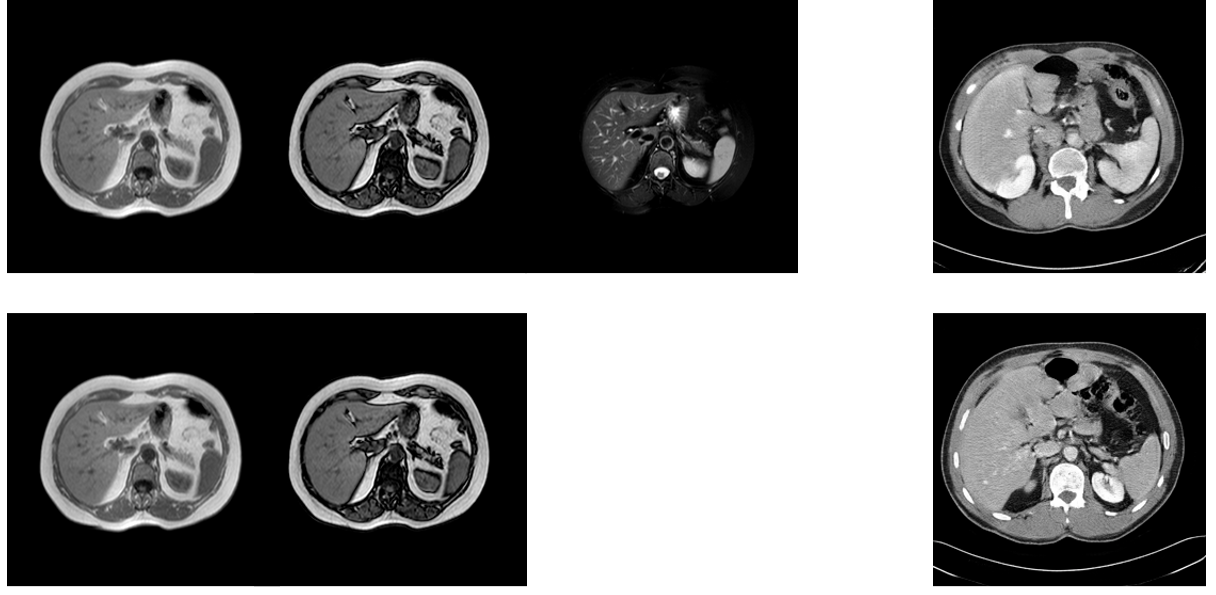}
		\caption{Top: CT slice generated with \emph{Triple} model; bottom: slice generated from the same MRI slice with \emph{In-n-out} model.}\label{fig:sliceJoin-joinT2}
	\end{subfigure}%
	\hspace*{\fill}
	\begin{subfigure}{0.48\columnwidth}
		\centering
		\includegraphics[width=\columnwidth]{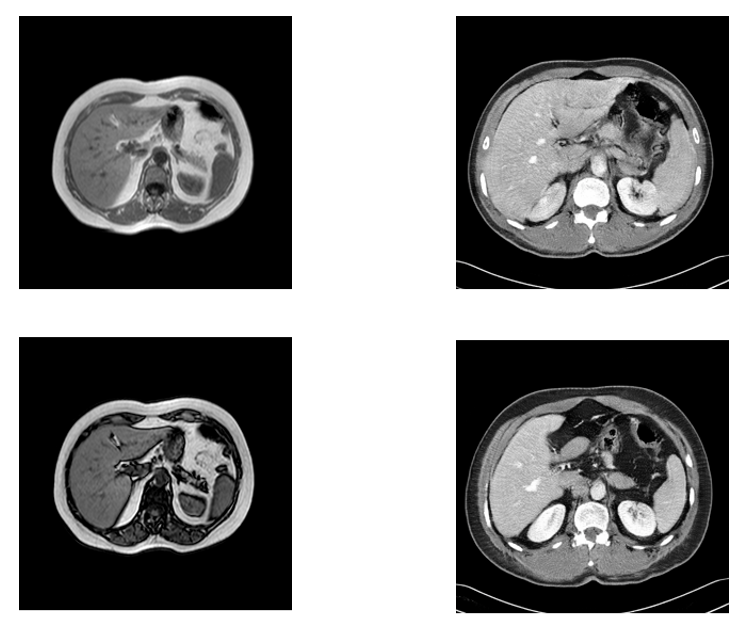}
		\caption{Top: CT slice generated with \emph{All-in} model from a T1-in-phase slice; bottom: slice generated from the same MRI slice, but in T1-out-of-phase mode, with \emph{All-in} model.}\label{fig:sliceFull}
	\end{subfigure}

	\begin{subfigure}{0.48\columnwidth}
		\centering
		\includegraphics[width=\columnwidth]{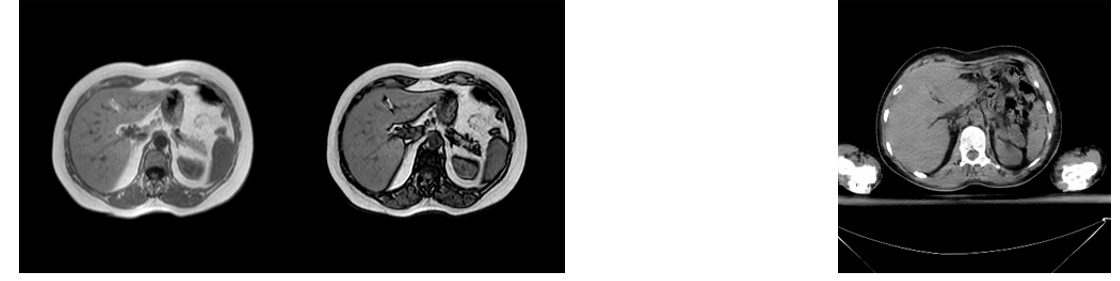}
		\caption{CT slice generated with \emph{In-n-out HUM} model from a T1-in-phase slice.}\label{fig:sliceHumSame}
	\end{subfigure}%
	\hspace*{\fill}
	\begin{subfigure}{0.5\columnwidth}
		\centering
		\includegraphics[width=\columnwidth]{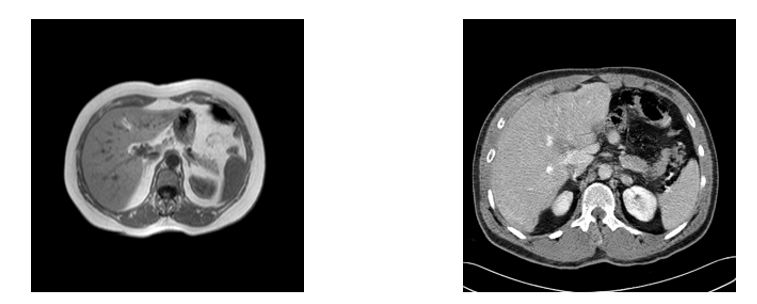}
		\caption{CT slice generated with \emph{In-phase} model from a T1-in-phase slice}\label{fig:sliceInPhase}
	\end{subfigure}

	\centering
	\begin{subfigure}{0.48\columnwidth}
		\centering
		\includegraphics[width=\columnwidth]{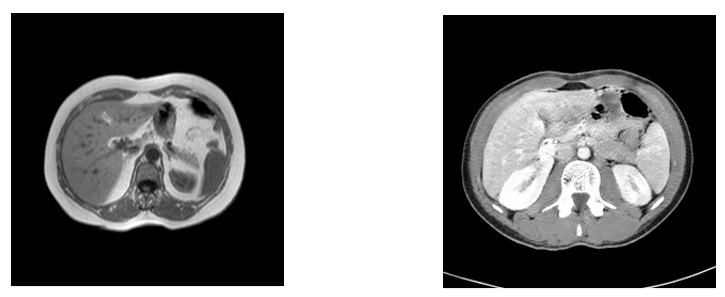}
		\caption{CT slice generated with \emph{In-phase+} model from a T1-in-phase slice.}\label{fig:sliceExtSame}
	\end{subfigure}
    \caption{Examples of synthetic slices from the different models; in each subfigure, the input is on the left, the output on the right.}\label{fig:samples}

\end{figure}

In the subsequent figures, we aim to discuss the tendencies exhibited by the various models from a qualitative perspective, comparing their output on the same MRI slice used in the different modalities available to produce the required outputs.
Beginning with fig.~\ref{fig:sliceJoin-joinT2}, it's evident that the model integrating in-phase and out-of-phase inputs delivers superior images compared to the \emph{Triple} model encompassing both T1 and T2 modalities.
This reaffirms the evidence, already clear from the distribution analyses, that T2-weighted images do not substantially contribute much to generating realistic CT scans when paired with the T1-weighted images.
Observing fig.~\ref{fig:sliceHumSame}, it can be seen that the model leveraging in-phase and out-of-phase T1 inputs but producing non-contrast-enhanced outputs yields results strikingly similar in organ shape to its counterpart generating contrast-enhanced outputs.
Fig.~\ref{fig:sliceFull} delineates the in-phase and out-of-phase images processed as distinct entities by the single-input CycleGAN model with the \emph{All-in} architecture.
Here, the in-phase image seems marginally superior to its out-of-phase counterpart, although, under ideal circumstances, both should yield analogous CT images.
This proves the model cannot yield reliable results as different MRI modalities produce different outputs.
It has to be noted that it would almost be impossible to get the same results, but here, it is clear how the generated synthetic anatomical structures are not consistent at all.
Furthermore, fig.~\ref{fig:sliceInPhase} demonstrates that the CycleGAN model, when trained solely on in-phase T1 images, produces outcomes akin to the in-phase image from the model trained on both in-phase and out-of-phase images independently.
This mitigates the inconsistency of the \emph{All-in} model, as with in-phase T1 images, its output is paired with other specialized models trained on that.
Lastly, delving into the performance of the model trained with multiple MRI modalities, fig.~\ref{fig:sliceExtSame} reveals an anomaly where it produces a fictitious kidney, signifying that the inclusion of images from such radically different sources can be problematic as the increment in the sample variability seems to lead to a more difficult distribution to learn for the model, which resorts to the creation of unrealistic structures, but the models seem to be able to learn how some structures are supposed to look (in this case, slices usually contain both kidneys).

These are only some examples, but the choice has been made on an exemplificative slice, capable of showing the most relevant differences between the models and performing comparisons that highlight the strengths and weaknesses of the models. 
The considerations taken here are generally backed up by the quantitative results that provide a clear ranking of models according to the fidelity of the distribution learnt with respect to the real one from the dataset.
From these examples, it is possible to infer that the quantitative analysis leads to a realistic direction in the model's evaluation.
This considered, the best models seem to be the multimodal ones, especially the ones trained on the Automi dataset and thus producing non-contrast-enhanced CT scans, which is expected as the use of contrast agents might introduce an additional layer of complexity to the problem. 

\subsubsection*{Blind evaluation}\label{ch:user_study}
A user study has been conducted in collaboration with Humanitas Research Hospital to evaluate the quality of the results further.
This partnership was part of the AUTOMI project and aimed to gauge how medical professionals perceive the generated CT scans compared to actual images.
Four different surveys were administered to the radiation oncologists, each about a different model:
\begin{itemize}
	\item Survey 1 for the \emph{In-n-out} model;
	\item Survey 2 for the \emph{All-in} model;
	\item Survey 3 for the \emph{Triple} model;
	\item Survey 4 for the \emph{In-n-out HUM} model.
\end{itemize}

Each survey presented the participants with 20 scans: 10 real and 10 synthetic.
As depicted in figure~\ref{fig:survey2}, the medical personnel were prompted for each scan to categorize it as real, synthetic or indeterminable.
They were also invited to state the rationale behind their decisions, potentially pointing out specific features or organs that influenced their judgment.
This approach helped us discern potential pitfalls in our models and furnished insights into the overall quality of the generated scans.

In table~\ref{table:user_study_table}, we provide a breakdown of the percentage of correct guesses across the different surveys.
These percentages are delineated for the entire set of scans and separated into real-only and synthetic-only groupings.
Notably, a percentage closer to 50\% indicates a desirable outcome, suggesting that the medical professionals found little to no discernible distinction between real and synthetic scans. The users did not know the number of real and synthetic slices in advance. 

\begin{table}
	\centering
	\begin{tabular}{|p{6em} | c | c | c |}
		\hline
		                  & \textbf{Full scan set} & \textbf{Real scans only} & \textbf{Synthetic scans only} \\
		\hline \hline
		\textbf{Survey 1} & 51.8\%                 & 49.0\%                   & 51.6\%                        \\
		\textbf{Survey 2} & 59.9\%                 & 66.7\%                   & 53.5\%                        \\
		\textbf{Survey 3} & 57.5\%                 & 59.3\%                   & 55.6\%                        \\
		\textbf{Survey 4} & 52.3\%                 & 55.4\%                   & 47.3\%                        \\
		\hline
	\end{tabular}
	\caption{Collective percentage scores of correctly classified scans in the user study by MDs.}
	\label{table:user_study_table}
\end{table}

The medical practitioner's feedback resonated with the findings of the quantitative evaluations.
Specifically, the \textbf{In-n-out} and\emph{In-n-out HUM} models surfaced as the top performers, with the doctors correctly identifying 51.8\% and 52.3\% of the scans, respectively.
While not leading the pack, the other two models still exhibited commendable results; the \textbf{Triple} model garnered a 57.5\% correct identification rate, and the \emph{All-in} model registered a score of 59.9\%.

For context, survey 1 received feedback from eleven doctors, survey 2 from eight doctors, and surveys 3 and 4 had six doctors each.

\begin{figure}
	\begin{subfigure}{0.3\columnwidth}
		\centering
		\includegraphics[width=\columnwidth]{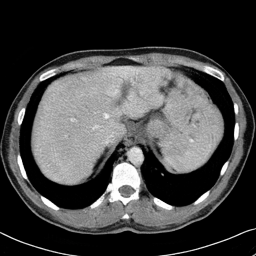}
		\caption{Synthetic CT showing an evident costal asymmetry.}\label{fig:asimCost}
	\end{subfigure}%
	\hspace*{\fill}
	\begin{subfigure}{0.3\columnwidth}
		\centering
		\includegraphics[width=\columnwidth]{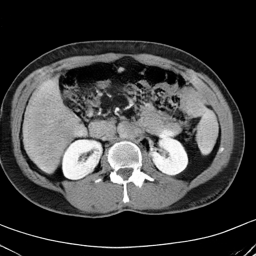}
		\caption{Synthetic CT with hyperdense and artificial kidneys.}\label{fig:reniArtefatti}
	\end{subfigure}%
	\hspace*{\fill}
	\begin{subfigure}{0.3\columnwidth}
		\centering
		\includegraphics[width=\columnwidth]{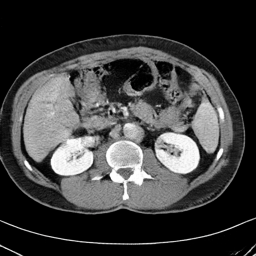}
		\caption{Synthetic CT with unclear delimitation plans between muscular structures.}\label{fig:clivaggio}	
	\end{subfigure}
	
	\begin{subfigure}{0.3\columnwidth}
		\centering
		\includegraphics[width=\columnwidth]{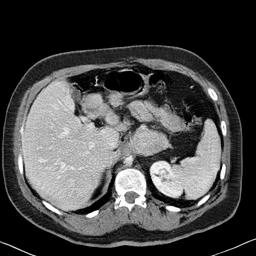}
		\caption{Synthetic CT with an unrealistic morphology of the spleen.}\label{fig:milza}
	\end{subfigure}%
	\hspace*{\fill}
	\begin{subfigure}{0.3\columnwidth}
		\centering
		\includegraphics[width=\columnwidth]{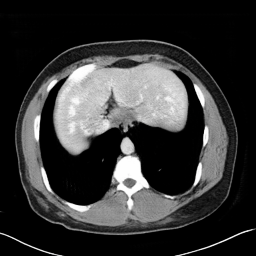}
		\caption{Synthetic CT with an unrealistic morphology of the liver.}\label{fig:badLiver}
	\end{subfigure}%
	\hspace*{\fill}
	\begin{subfigure}{0.3\columnwidth}
		\centering
		\includegraphics[width=\columnwidth]{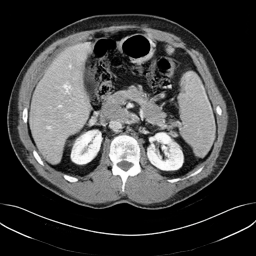}
		\caption{Synthetic CT where the stomach is in an unrealistic position.}\label{fig:stomach}
	\end{subfigure}

	\hspace*{\fill}
	\begin{subfigure}{0.3\columnwidth}
		\centering\centering
		\includegraphics[width=\columnwidth]{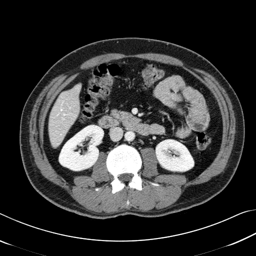}
		\caption{Real CT described as difficult to evaluate due to the not ideal contrast.}\label{fig:badRange}
	\end{subfigure}%
	\hspace*{\fill}
	\begin{subfigure}{0.3\columnwidth}
		\centering\centering
		\includegraphics[width=\columnwidth]{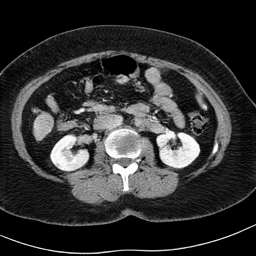}
		\caption{Synthetic CT described as anatomically accurate.}\label{fig:good}
	\end{subfigure}
	\hspace*{\fill}

	\caption{Examples of slices with a summary of the comments from the physicians in the survey.}

\end{figure}

Regarding the comments made by doctors about the elements driving their decisions, we present some examples in the following figure;
fig.~\ref{fig:asimCost} shows how a common problem for generated scans is the asymmetry in the ribs of the patient;
another recurring challenge pertained to the manifestation of hyperdense or artificial kidneys, as evident in figure~\ref{fig:reniArtefatti}.

Other common observations include:
\begin{itemize}
    \item Costal asimmetry~\ref{fig:asimCost}.
	\item Unclear delineation between muscular structures, as evident in figure~\ref{fig:clivaggio}.
	\item The spleen that appeared either poorly defined or morphologically abnormal, demonstrated in figure~\ref{fig:milza}.
	\item Stomach in an unrealistic position~\ref{fig:stomach}.
	\item Vaguely defined blood vessels.
	\item An imprecisely represented liver, highlighted in figure~\ref{fig:badLiver}.
\end{itemize}

Interestingly, a few physicians also articulated their rationale for determining specific scans as real.
For instance, fig.~\ref{fig:good}, which is synthetic, was mistakenly identified as real, and it even received a commendation from an MD for its \emph{high definition, quality, and structural density grading}.

fig.~\ref{fig:badRange} presents a scan critiqued by a medical professional due to its difficulty in assessment within the provided window width.
Producing CT scans spanning multiple HU ranges can be instrumental in accentuating specific anatomical structures.
This methodology will be taken under advisement for prospective enhancements of this research and is expounded upon in the next section.

The Chi-squared test has been employed as a statistical method to assess the distribution of the responses in two main scenarios: comparing two distinct surveys and comparing real versus synthetic images.
For the latter, analyses were conducted on images from a single survey and the complete collection of scans across all four surveys.
This test aimed to understand if there was a consistent distribution of the doctors' responses across different survey comparisons.
In this analysis, the survey responses have been treated either as binomial or multinomial variables.
In the binomial categorization, a value of 1 indicates a scan is classified as real by the physician, whereas a 0 signifies it is classified as synthetic.
In the multinomial classification, an additional value of 2 is employed to denote scans that were marked as \emph{indeterminable} by the respondents.

For the Chi-squared test, the null hypothesis \( H_0 \) posits that there is a dependence between the two series compared, implying that they originate from the same distribution.

\begin{table}
	\centering
	\begin{tabularx}{0.6\textwidth}{| l | c | c |}
		\hline
		                                        & \textbf{Binomial} & \textbf{Multinomial} \\
		\hline \hline
		\textbf{Survey 1: real vs synthetic}    & 0.933             & 0.478                \\
		\textbf{Survey 2: real vs synthetic}    & 0.027             & 0.000                \\
		\textbf{Survey 3: real vs synthetic}    & 0.178             & 0.175                \\
		\textbf{Survey 4: real vs synthetic}    & 0.930             & 1.000                \\
		\textbf{All surveys: real vs synthetic} & 0.045             & 0.057                \\
		\textbf{Survey 1 vs Survey 2}           & 0.185             & 0.022                \\
		\textbf{Survey 1 vs Survey 3}           & 0.625             & 0.916                \\
		\textbf{Survey 1 vs Survey 4}           & 0.387             & 1.000                \\
		\textbf{Survey 2 vs Survey 3}           & 0.582             & 0.079                \\
		\textbf{Survey 2 vs Survey 4}           & 0.044             & 0.040                \\
		\textbf{Survey 3 vs Survey 4}           & 0.848             & 0.892                \\
		\hline
	\end{tabularx}
	\caption{P-values of the Chi-squared tests.}\label{table:p-values}
\end{table}

Table~\ref{table:p-values} delineates the resulting p-values from our comparisons, showing both the binomial and multinomial versions of the analysis.
A p-value greater than 0.05 has been chosen to reject the null hypothesis \( H_0 \). Consequently, we lack sufficient evidence to claim that the two compared series emanate from different distributions.
These observations reveal that the distributions of real and synthetic scans in Surveys 1 and 4 are not different.
Similarly, for Survey 3, given that all the resulting p-values surpass 0.05, the null hypothesis stands.
Once again, when comparing Surveys 1, 3, and 4 directly, their p-values do not provide enough grounds to determine that they derive from different distributions.
The outlier in our evaluations is Survey 2, which exhibits a p-value below 0.05. Hence, we can reject the null hypothesis, implying a discernible difference between its real and synthetic scan distributions.

In general, the most interesting conclusion related to this part of the work on synthetic image generation is that the models are indeed capable of synthesizing images that are at least very difficult to distinguish from real ones, and even expert physicians are not able to distinguish them consistently,
proving the capabilities of the models and the potential of this approach to tackle challenges such as data augmentation and data generation for medical purposes. 

%% file: sections/conclusions.tex

\section{Discussion}
Our experiments with various CycleGAN architectures revealed the versatility of these models in generating synthetic CT scans from corresponding MRI scans.
Notably, we discerned the superior performance of MRI T1-weighted scans over their T2 counterparts.
The multi-input methodologies outperformed other configurations, particularly those leveraging in-phase and out-of-phase T1-weighted MRIs.
Nevertheless, single-input models, such as the MRI in-phase only, still exhibited commendable efficacy.
Remarkably, our models adeptly generated both contrast-enhanced and non-contrast-enhanced CTs, achieving respectable results in each modality.

A non-trivial challenge encountered was evaluating generated images from unpaired data sets.
Our assessment matrix encompassed metrics like FID and KL divergence, enabling us to gauge the efficacy of our models with precision.
These quantitative metrics, juxtaposed with our qualitative evaluations, yielded insights into the performance landscape, highlighting both the strengths and limitations of the trained models, proving that the models which were able to learn the distribution from the original sets better were indeed the
multimodal ones, particularly the so-called \emph{In-n-out HUM} trained with MRI from the Chaos dataset and non-contrast-enhanced CT scans from the  Automi dataset.
Through a spectral analysis, we delved deeper into understanding the quality of the synthesized images, thus confirming that the capabilities of the various models indeed emerged with the purely quantitative analysis of pixel distribution.

Qualitative evaluation is, however, an essential part of the work, as, since the field is particularly tricky in terms of structures, proportions, textures, relative positions and so on, it is of paramount importance to produce something which, more than on a statistical point of view, is
accurate from a medical standpoint.
Working with physicians in a blind trial to test if they could distinguish real from synthetic images, it has emerged that the accuracy and quality of the generated images are relatively high, proven by the fact that they could not classify them as real or synthetic consistently.
This, however, poses some approach problems that make the test not wholly reliable, as this is not a task that a physician would typically do in their daily routine or ever.
First of all, the images were not shown in a clinical context but in a more relaxed environment, which is not the same as where they would typically work.
This environment did not allow for adjusting the image's contrast, which is very important for radiation oncologists, especially when analyzing CT scans; they can play with it to enhance the visibility of certain structures or areas.
Doing so, especially on uniform structures such as the bones, it is possible to see a non-uniform texture in the synthetic images, which would make it easier to distinguish them from the real ones.
Moreover, a limitation of this approach is the use of a 2D-only approach and the generation of CT slice by slice without information on the $3_{rd}$ dimension.
Usually, a doctor would work on a whole volume, which is extremely difficult to recreate.

Nonetheless, the approach presented can be considered extremely promising as the statistical metric shows interesting performances, and the qualitative evaluation is still very positive, even if not wholly reliable.
In synthesis, our holistic evaluation approach—merging quantitative analyses, spectral evaluations, and real-world testing—has furnished robust insights into the promising behaviour of the models trained.
These findings validate our approaches' efficacy and pave the way for future advancements in this realm.